# Photonic crystal-Microring resonators for tunable delay lines

Sumanyu Chauhan and Roza Letizia

**Abstract**   In this paper, a new design of optical delay line based on coupled ring resonators in photonic crystals is proposed and analyzed by means of numerical simulation in CST Microwave Studio. The performance of the proposed photonic crystal ring resonator (PhCRR) based coupled resonant optical waveguide (CROW) is compared to:
- point-defect cavity based coupled cavity waveguides (CCW) in photonic crystal.
- microring resonator (MRR) based CROW.

The suggested design addresses compactness issues of the MRR CROW while offering normalized delay values better than CCW and comparable to MRR CROW. The PhCRR CROW also enhances design flexibility as the PhCRRs support multiple modes and thereby supporting multi-channel transfer/delay application on a single device.

------------------------------------------

R. Letizia
Engineering Department, Lancaster University, Lancaster - LA1 4YR, United Kingdom

S. Chauhan
Engineering Department (Alumni), Lancaster University, Lancaster - LA1 4YR, United Kingdom

# 1 Introduction

Improving optical buffers is one of the most challenging issues in the realization of optical delay lines. Slow-light optical devices have been intensively studied because of their properties for handling the flow of light. Among the variety of features connected to the concept of slow light, the ability of controlling the optical group velocity and/or the group delay of the light can find applications in the realization of variable delay lines and devices for optical storage and buffering [1].

For control of optical pulse in the slow-light regime coupled optical resonators are typically used. Some examples of optical resonators are: a ring, a disc, a toroid, a sphere, an annular Bragg grating, a defect in a photonic crystal lattice and a Fabry-Perot cavity [2]. Over the years (microring resonator) MRR has emerged as a component for integration into photonic circuits because unlike MEMS it does not have any moving part thus a longer life period. To control the group velocity of light, coupled MRRs have been effectively shown as delay lines [3].

It has been reported that as the diameter of the MRR is reduced below 5μm, propagation and bending losses increase exponentially, negatively affecting performance [4]. This brings about the need to find an alternative technology which provides reduction in size and provides opportunity for improving performance in slow-light regime. It is well known that light trapped inside PBG material like photonic crystal (PhC) can be guided through tight bends because whatever modes light can diffract are removed by the PBG. Optical resonators can be created in photonic crystal by creating point and line defects. Point and line defects in PhC can be coupled to gain control over group velocity of light [5] [6].

PhC based delay lines have emerged as an attractive candidate for such applications, as the dimensions of the stored bit is comparable to the wavelength [7]. Other advantages of integrating PhC in microchips include: reduction in back scattering along the circuit path [8], unlike electronic chips heat generation and dissipation are no issue in 3D optical microchips [9].

Coupled cavity waveguides on a square lattice with point defects have shown velocity reduction up to $0.4c$ [10]. Apart from point and line defects another type of defect inspired from the microring resonator had been implemented in photonic crystal. Ring resonator defects in PhC have much higher Q-factor than point/line defects [11]. In this paper, we use coupled circular ring resonators [12] in PhC to form a delay line.

In this paper, following the introduction, the formulas used for calculating performance metrics of MRR CROW and PhCRR CROW are provided in Section 2. Section 3 provides the simulation results for a MRR CROW and PhCRR CROW obtained in CST Microwave Studio. Finally, conclusions are given.

# 2 Analysis

Microring resonator (MRR) consists of an optical waveguide looped back on itself, when the number of wavelengths in the circular resonator are in whole multiples of the optical path, the energy in the structure resonates. Microring resonators play an important role in silicon photonics and have been heavily researched into. MRRs can be coupled to form CROWs which can be used as optical delay lines. The theory behind working of MRRs and its various configurations such as MRR CROWs has been explained in detail by Bogaerts, et al. [13]. The maximum group velocity at the center of CROW transmission band can be calculated using the formula provided in [14]

Mechanisms proposed in the past for optical waveguiding include: total internal reflection, Bragg waveguiding and waveguiding based on coupling of optical resonators or coupled resonator optical waveguide (CROW) [15] [16]. Some of the possible ways realizations of CROW based waveguides are:

evanescent-field coupling between the high-Q modes of individual microdisk cavities in MRR CROW and evanescent-field coupling between the individual resonators (point defect cavities) in [17] [18] embedded in a two-dimensional periodic structure such as 2D photonic crystal [19] [20]. The coupling in case of CROWs is due to the evanescent Bloch waves.

In [21], Yariv et el., employed a formalism similar to the tight-binding method in solid-state physics, and obtain the relations for the dispersion and the group velocity of the photonic band of the CROW's and explained that they are solely characterized by coupling factor (k). For development of this formalism for CROWs, Yariv et al. assumed that sufficiently large separation is present between the individual resonators so that the resonators are weakly coupled. Consequently, it is expected that the eigenmode of the electromagnetic field in such a coupled-resonator waveguide will remain essentially the same as the high-Q mode in a single resonator.

In this paper, we propose a new design for delay lines in PhC. Instead of coupling point defects in PhC, we create a waveguide by coupling ring-resonators in PhC. Ring-resonators in PhC exhibit much higher-Q [11] than point defects. Therefore, the higher-Q of ring-resonators can be utilized in creating delay-lines by coupling them to form a CROW in PhC, similar to the MRR CROW. **The coupling in PhCRR CROW, essentially (just like in case of CCWs) occurs due to coupling of evanescent Bloch waves. The ring-resonators in PhCRR CROWs are sufficiently separated such that they are weakly coupled and the EM-field in the PhCRR CROW remains essentially the same as EM-field in a single resonator. Since, the undelaying assumptions in the case of PhCRR CROW and its behavior (observed from CST simulations) are same as that of CCW [21], we anticipate that the same formalism can be applied to PhCRR CROW. Therefore, formulas used for circular PhCRR CROW are given as** [10] [21] [22] **:**

$$v_g = \pm R\Omega \sqrt{\kappa^2 - \left(\frac{\omega}{\Omega} - 1\right)^2} \qquad (1)$$

Where:
$v_g$ is the group velocity of the EM-mode.
$R$ is the cavity spacing.
$\kappa$ is the coupling coefficient.
$\omega$ is the operating frequency.
$\Omega$ is the center frequency.
$\kappa = \triangle\omega/2$

At center frequency ($\Omega$), $v_g$ becomes:

$v_g = \kappa R\, \Omega$

### 3 Simulation results

**3.1 MRR CROW**

The MRR in an all-pass filter configuration is shown in figure 1 was simulated in CST Microwave Studio. The design parameters were similar to that of the previous work reported in [23], to validate the results obtained in terms of the resonant peaks and coupling coefficients as given in table 1.

Figure 1, shows schematic of the MRR simulated and placement of field probes within the simulated geometry. Input probe 1 and output probe 1 detect the input and output fields respectively, while 1, 2 and 3 detect the resonant fields in the ring resonator. The 2D structure was simulated in the x-y plane, where the perpendicular direction along the z-axis is considered infinite. The frequency range simulated was 180-220 THz, the first resonant mode within this band for the given parameters has a modal number of m = 25. This means that the resonator can accommodate 24 other resonant modes on the lower side of the frequency range and many other modes on the higher side of the frequency range (depending on the frequency tolerance of the material of MRR). In figure 2, the graph shows the FFT of E-field, as computed in the ring probe 1. It shows resonant frequencies and corresponding modal number within simulated frequency range.

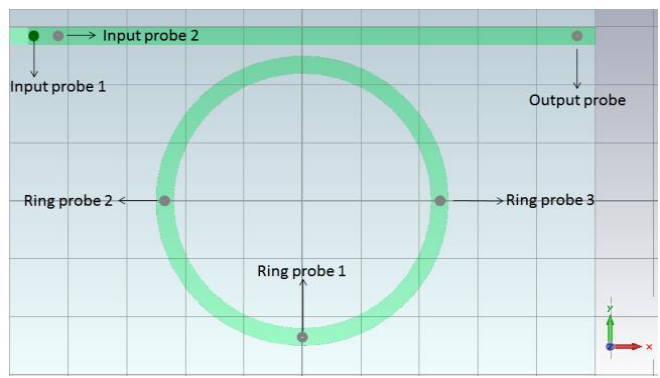

Fig. 1 Schematic of MRR in APF configuration with placement of field probes.

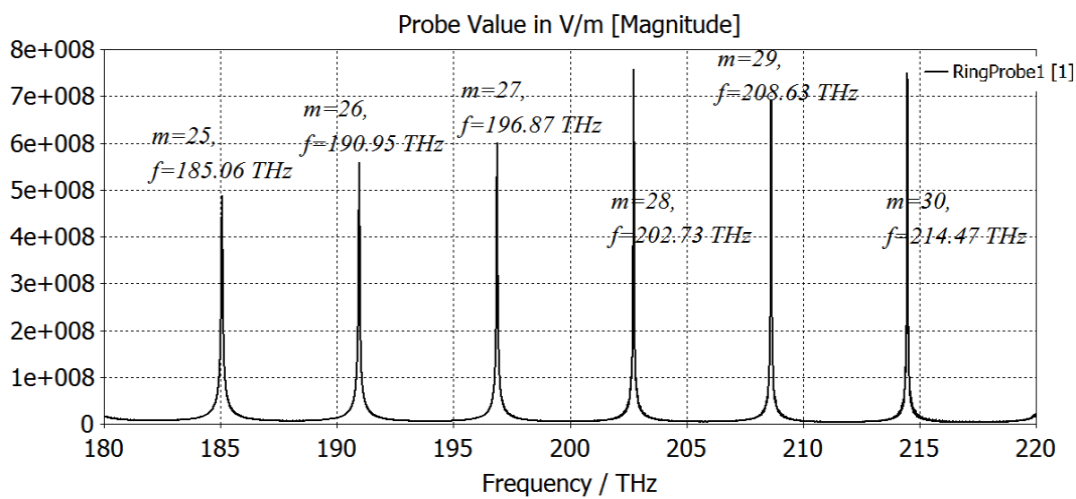

Fig 2. Resonant frequencies for the MRR within the simulated frequency range.

Figure 3 shows four conventional ring-resonators coupled to each other, the placement of field-probes in the ports and in the ring coupled to the bus waveguide. The gap between two adjacent rings is same as the gap between the ring and waveguide. The structure was simulated with periodic boundary conditions along the z-axis and open boundary condition in the x-y plane. The structure was discretized using hexahedron mesh and simulated within a frequency range of 180-220 THz. The maximum duration of the simulation was set to 200 pulse widths (200 times the width of excitation signal).

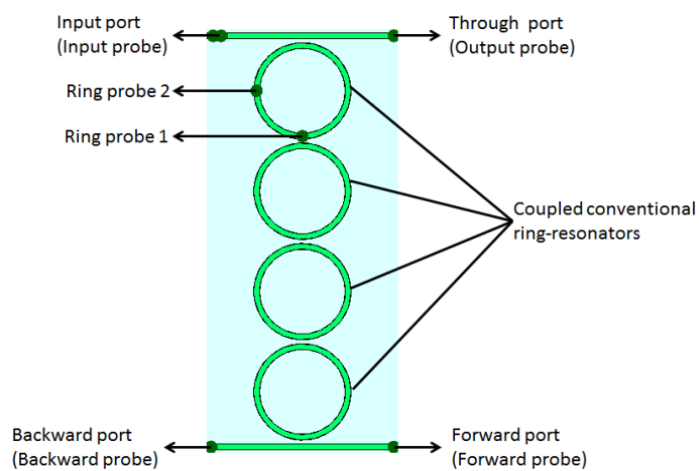

Fig. 3 Schematic of MRR CROW with the placement of field probes.

Figure 4 shows the frequencies dropped in the forward port of the MRR based CROW. It was observed that for the odd modal numbers (m=25, 27 and 29), higher amplitude of the field is transmitted, this is due to the unidirectional circulating nature of the resonant modes with odd modal number.

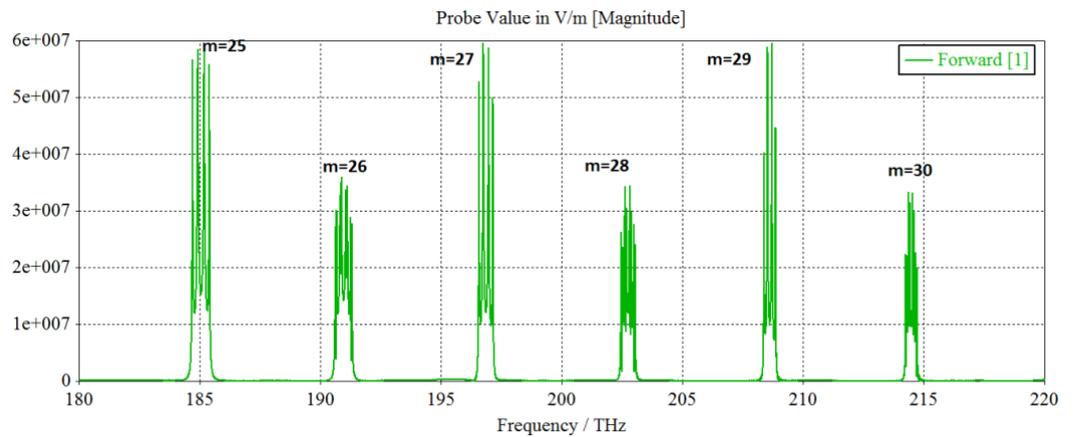

**Fig. 4** Transmission characteristics for the different modal numbers in the forward port of MRR CROW.

Figure 5 (a) shows the transmission characteristics for odd modal number (m=29) in the forward port of MRR based CROW. There are four resonant peaks for the transmission bandwidth. The four peaks are marked as 1, 2, 4 and 5. The four peaks occur as a consequence of four coupled ring-resonators where each ring-resonator adds negative or positive frequency shift around the central frequency of the isolated ring resonator. Figure 5 (b) shows the transmission characteristic for even modal number (m=30) in the forward port of MRR based CROW. There are four resonant peaks for the transmission bandwidth. The four peaks are marked as 1, 2, 4 and 5. The four peaks occur as a consequence of four coupled ring-resonators where each ring-resonator adds a shift in frequency around the central frequency of the transmission bandwidth. The two blue rings for peak 1 represent the two degenerate modes for the first resonant peak. The resonance splitting occurs as a consequence of counter-directional coupling. Same holds true for the resonance-splitting in peak 2, 4 and 5.

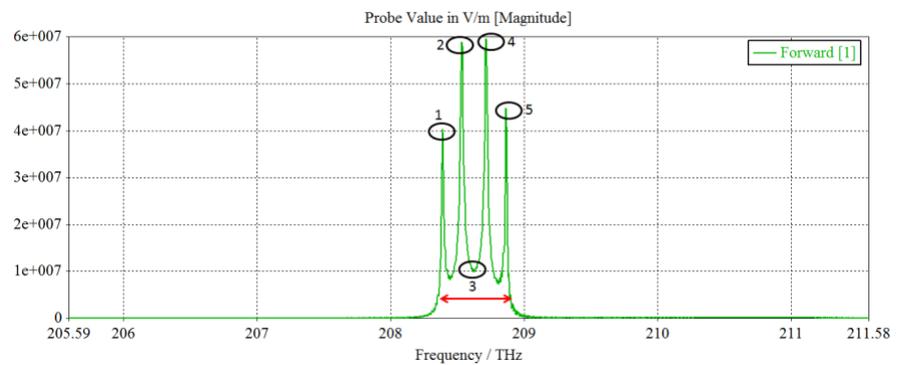

**Fig 5 (a)** Transmission band for m=29

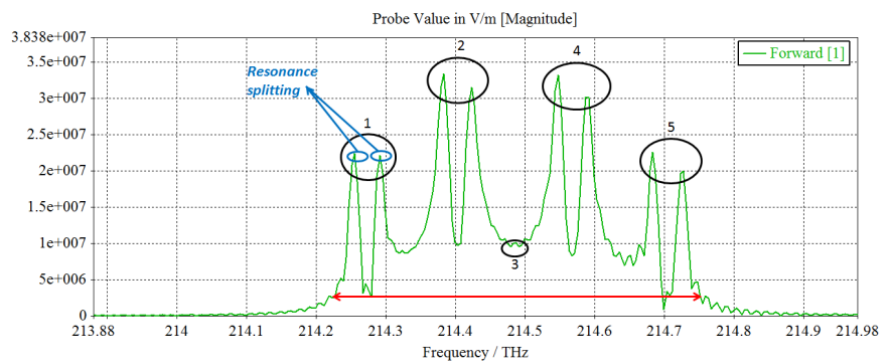

**Fig. 5 (b)** transmission band for m=30

The normalized group velocity of a resonant mode in MRR CROW are shown in table 1. These were determined using equation from [14].

Table 1 Modal number (m), resonant frequency (F$_{res}$) and wavelength (λ$_{res}$), effective refractive index (n$_{eff}$), coupling coefficient (k) and normalized group velocity values for the simulated MRR CROW.

| m  | F$_{res}$ (THz) | λ$_{res}$ (nm) | n$_{eff}$ | k      | Normalized Group velocity |
|----|-----------------|----------------|-----------|--------|---------------------------|
| 25 | 184.64          | 1624.783362    | 2.585923  | 0.06   | 0.01534                   |
| 26 | 190.56          | 1574.307305    | 2.605811  | 0.0471 | 0.01195                   |
| 27 | 196.52          | 1526.562182    | 2.623967  | 0.0389 | 0.00980                   |
| 28 | 202.51          | 1481.408326    | 2.640663  | 0.0311 | 0.00778                   |
| 29 | 208.47          | 1439.055979    | 2.656781  | 0.025  | 0.00622                   |
| 30 | 214.38          | 1399.384271    | 2.672627  | 0.0206 | 0.00509                   |

## 3.2 PhC ring resonator crow

Figure 6(a) shows the schematic of the PhCRR and figure 6(b) shows enlarged view of a portion of the PhCCR with the original rod position, marked by the blue circles, in the square lattice.

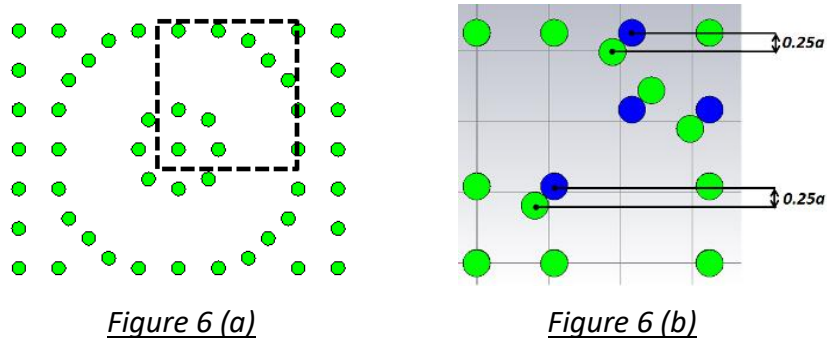

*Figure 6 (a)*           *Figure 6 (b)*

To validate the simulation results, initially the PhCRR was created with silicon rods placed in air in a square lattice with dimension (r= 0.175*a; a= 540; n$_{rod}$ = 3.47) was simulated and the resonances were verified with the results published Calo et al [12].

To improve the Q factor of the PhCRR the fill–factor was changed to 0.185*a and refractive index to n$_{rod}$ = 3.59 (gallium phosphide). The Preliminary analysis of the resonant modes was done by simulating a square lattice with an array of 15x15 dielectric rods placed in air. The cavity was excited by a plane wave using time-domain solver. The FFT of resonances detected is shown in the figure 7.

**Fig. 7**: FFT of resonances detected.

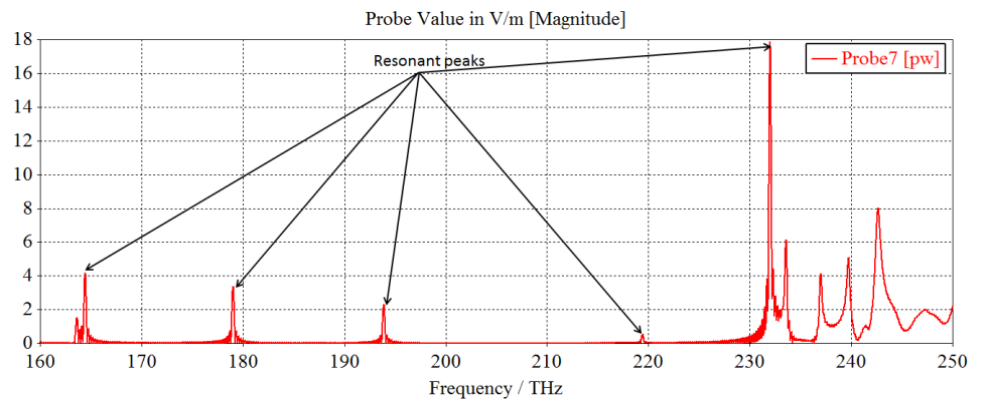

Figure 8 (a) shows schematic PhCRR CROW created by coupling circular ring resonator cavities on a PhC lattice (r= 0.185a ; a= 540; $n_{rod}$ = 3.59), for which resonant peaks are shown in figure 7. The cavity spacing (R) or the distance between the centers of the two cavities was taken as 3.240 µm. The PhCRR based CROW in figure 8 (a) was simulated in 2D environment with periodic boundary conditions along the z-axis and open boundary condition in the x-y plane. The structure was discretized using tetrahedral mesh and simulated using a frequency domain solver within a frequency range of 160-250 THz.

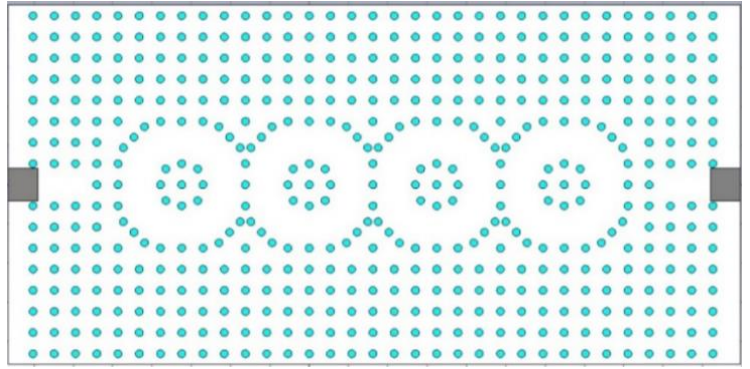

**Fig. 8 (a)** PhCRR CROW with cavity spacing as 3.240 µm

Figure 8(b) shows the variation of scattering parameters S (2, 1) for circular PhCRR CROW. The graph shows four transmission bands, since a single PhCRR (with lattice parameters as : r= 0.185a ; a= 540; $n_{rod}$ = 3.59) exhibits five resonant modes within 160-250 THz (as shown in figure 8), therefore five transmission bands are expected such that each transmission band in the PhCRR CROW has four phase shifted peaks. However, it can be noticed that there are only four transmission bands such that the first and the fourth transmissions bands have more than four phase shifted peaks each. This indicates that there is some interacting of hybrid modes in the first and fourth transmission bands. In figure 8 (b) , the second and third transmission bands have four phase shifted peaks each which means there is no mixing of modes, therefore these transmission bands will be used to closely analyze the slow-light behavior for the given structure.

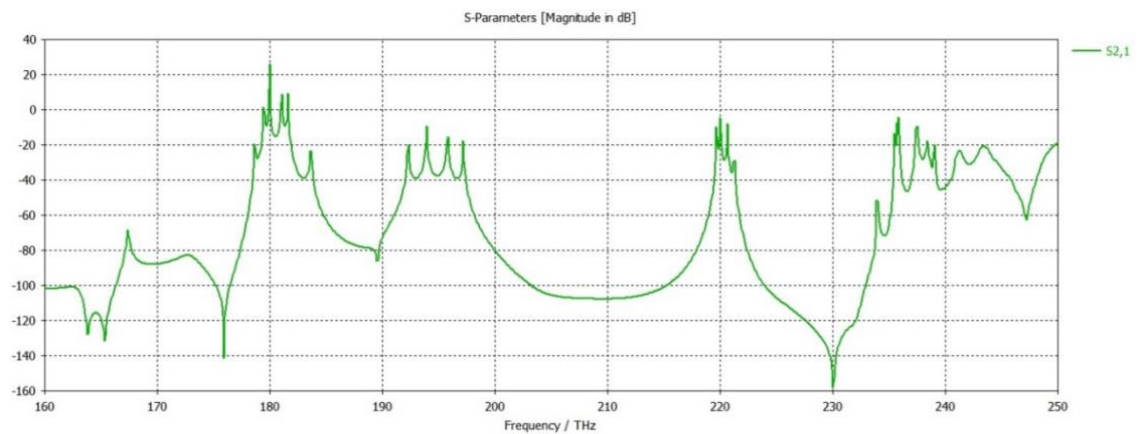

**Fig 8 (b)** Variation of S (2, 1) within 160-250 THz in a PhCRR CROW with cavity spacing as 3.240 µm.

Figure 8 (c) shows the variation of scattering parameters S (2, 1) for circular PhCRR CROW within frequency range of 190-200 THz. It should be noted that the central frequency of this transmission band (194.73 THZ) is also the resonant frequency of an isolated cavity (hexapole) in the PhCRR but due to change in the structure around the cavity, the center frequency hexapole in PhCRR differs slightly from the resonant frequency of isolated PhCRR. When many such cavities are coupled to form a PhCRR CROW then a frequency shift is added on either side of the central frequency due to each cavity [5]. In this case, four coupled cavities were simulated and as such four peaks can be seen in figure 8 (c). The 3db bandwidth of this transmission band was calculated as 4.92 THz in table 2 (given at the end).

**Fig 8 (c)** Variation of S (2, 1) within 190-200 THz range in a PhCRR CROW with cavity spacing as 3.240 µm. Transmission band centered at 194.73 THz.

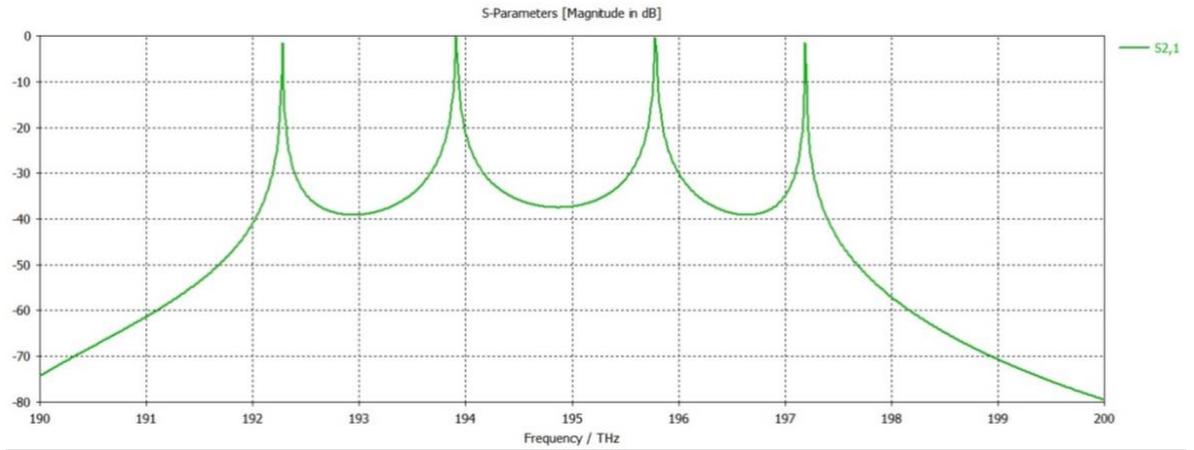

Similarly, figure 8 (d) shows the variation of scattering parameters S (2, 1) for circular PhCRR CROW within frequency range of 215-225 THz. It should be noted that the central frequency of this transmission band (220.48 THZ) in PhCRR CROW is also the also very close to the resonant frequency of an octupole in an isolated PhCRR cavity. The 3db transmission bandwidth for this band is 1.60 THZ as shown in table 2. Lower transmission bandwidth (for octupole as compared to hexapole). This property is again reflected in the coupling coefficient and normalized group velocity values shown in table 2. That is, octupole has lower coupling coefficient and lower normalized delay values compared to the hexapole.

**Fig 8 (d)** Variation of S (2,1) within 215-225 THz in a PhCRR CROW with cavity spacing as 3.240 µm. Transmission band centered at 220.48 THz.

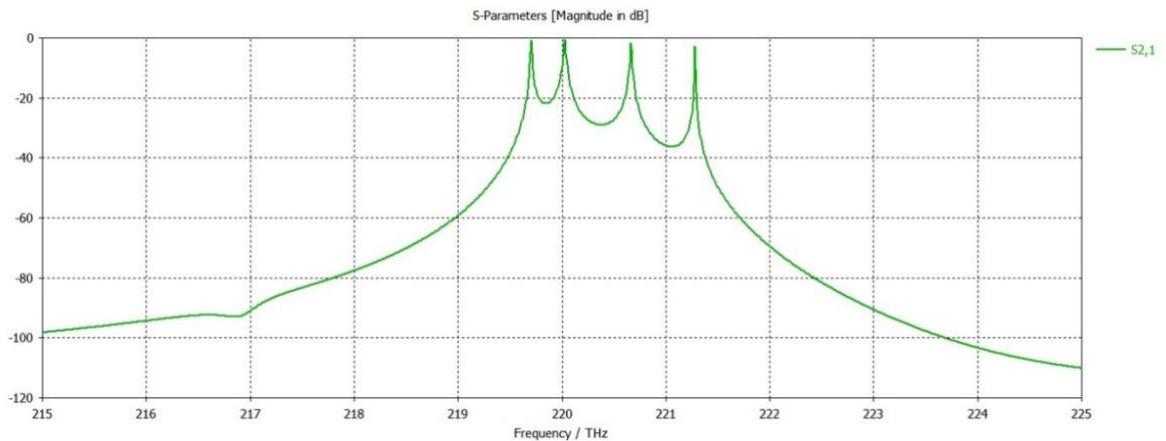

Figure 8 (e) and 8 (f) show mode profiles Hexapole and Octupole in the PhCRR CROW with cavity spacing of 3.240 µm.

**Fig 8(e)** Mode profile for heaxapole (192.28 THz) in PhCRR CROW with cavity spacing of 3.240 µm.

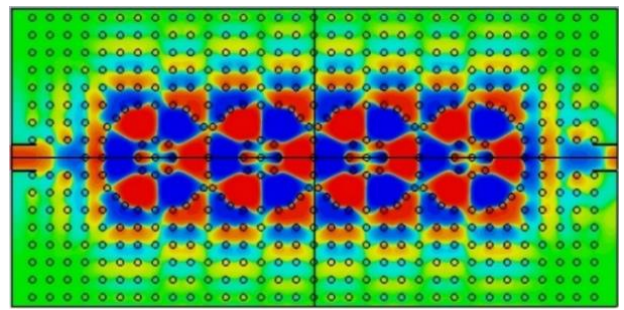

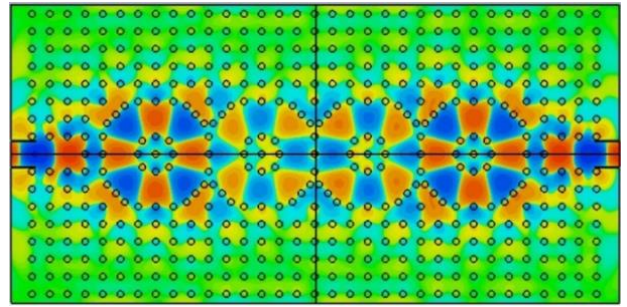

**Fig 8(f)** Mode profile for octupole (219.71 THz) in PhCRR CROW with cavity spacing of 3.240 μm.

To demonstrate to tunability of PhCRR CROW, the cavity spacing (R) was changed. Figure 9 (a) shows schematic of PhCRR CROW created by coupling circular ring resonator cavities on a PhC lattice (with lattice parameters as: r= 0.185a; a= 540; $n_{rod}$ = 3.59), for which resonant peaks are shown in figure 7. However, in this PhCRR CROW, the cavity spacing (R) or the distance between the centers of the two cavities was increased to as 4.320 μm. The PhCRR based CROW in figure 9 (a) was simulated in 2D environment with periodic boundary conditions along the z-axis and open boundary condition in the x-y plane. The structure was discretized using tetrahedral mesh and simulated using a frequency domain solver within a frequency range of 160-250 THz.

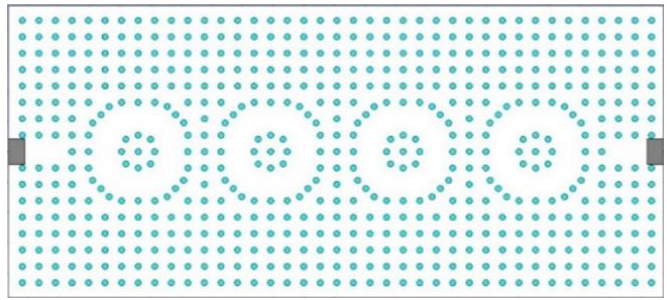

**Fig 9 (a)** PhCRR CROW with cavity spacing as 4.320 μm

Figure 9 (b) shows the variation of scattering parameters S (2, 1) for circular PhCRR CROW within frequency range of 193-196 THz. It should be noted that the central frequency of this transmission band (194.32 THZ) is also the resonant frequency of an isolated cavity (hexapole) in the PhCRR but due to the change in the structure around the cavity, the resonant frequency of hexapole in isolated PhCRR differs slightly from the center frequency of the transmission band for hexapole in PhCRR CROW.

The 3db bandwidth of this transmission band was calculated in table 2 as 0.44 THz; which is much lower than the hexapole in the PhcRR CROW will smaller cavity spacing (R=3.240). This happens because the field confinement improves in a PhCRR CROW with bigger separation between cavities. By contrasting schematics in figure 8 (a) and 9 (a) it can be observed that there are two extra rows of pillars separating the cavities in figure 9 (a).The extra row of pillars increase the confinement of the individual cavities.

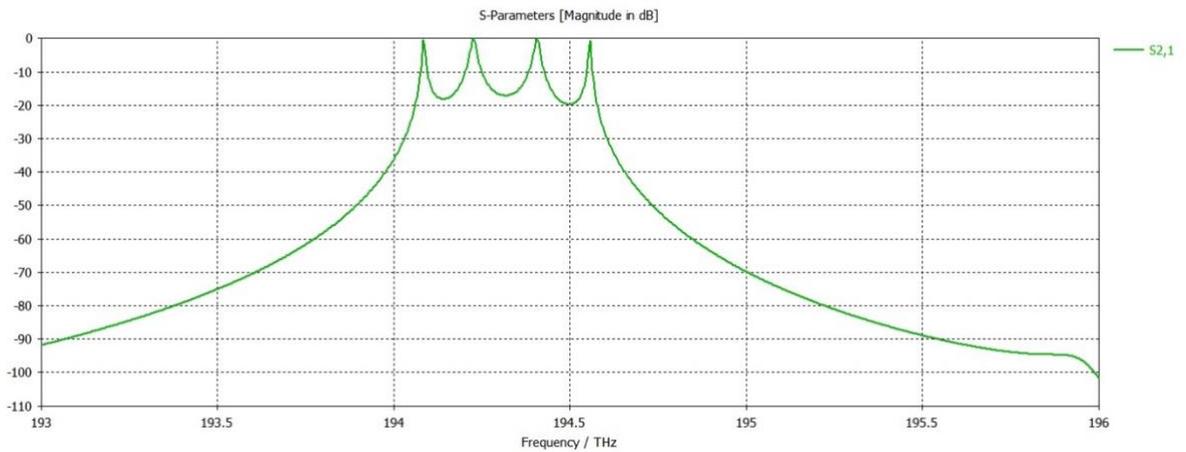

**Fig. 9 (b)** Variation of S (2, 1) within 193-196 THz range in a PhCRR CROW with cavity spacing as 4.320 µm. Transmission band centered at 194.32 THz

Similarly, figure 9 (c) shows the variation of scattering parameters S (2, 1) for circular PhCRR CROW within frequency range of 216-224 THz. It should be noted that the central frequency of this transmission band (219.85 THZ) is the resonant frequency of an isolated cavity (Octupole) in the PhCRR but due to slight variation in the structure around the cavity the center frequency varies slightly. The 3db bandwidth of this transmission band was calculated in table 2 as 0.48 THz which is again much lower in comparison to the bandwidth of the same mode (octupole) in PhCRR CROW with smaller cavity spacing (R=3.240 µm).

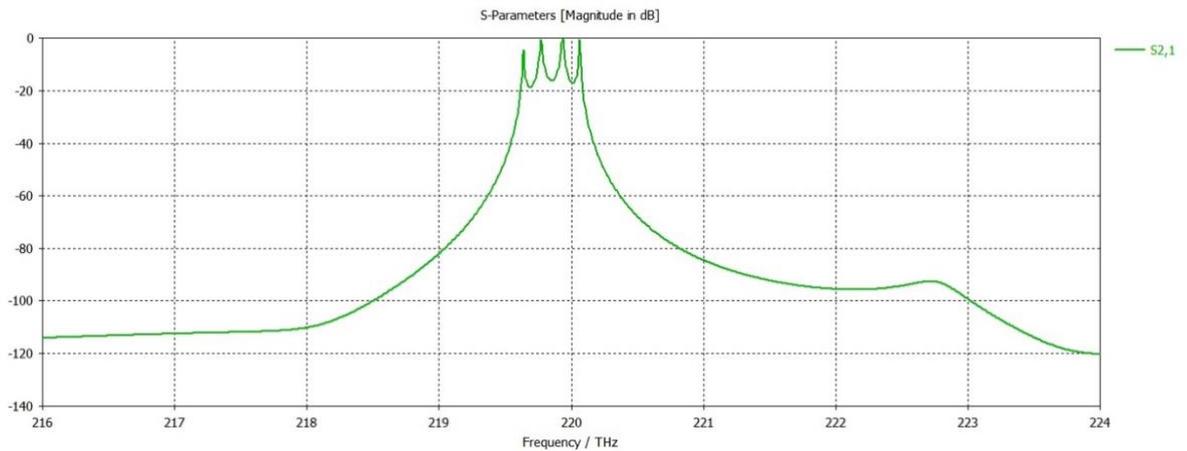

**Fig 9 (c)** Variation of S (2,1) within 216-224 THz in a PhCRR CROW with cavity spacing as 4.320 µm. Transmission band centered at 219.85 THz

Figure 9 (d) and 9 (e) show mode profiles of hexapole and octupole in the PhCRR CROW with cavity spacing of 4.320 µm.

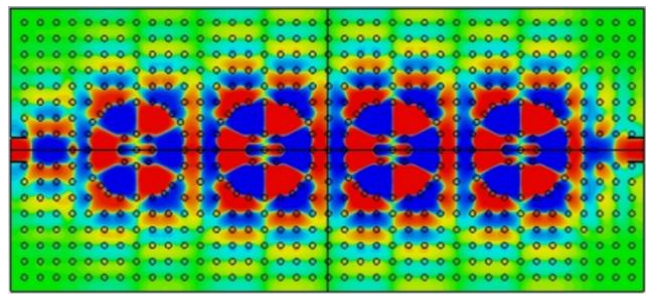

**Fig 9 (d)** Mode profile for heaxapole (194.32 THz) in PhCRR CROW with cavity spacing of 4.320 µm

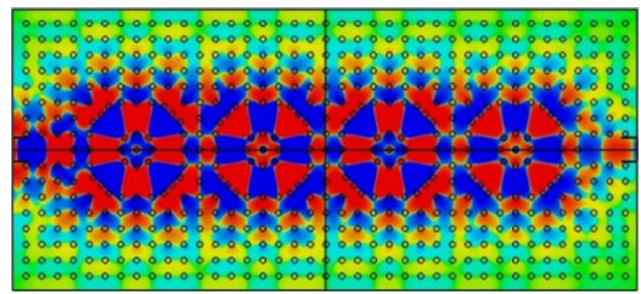

**Fig 9 (e)** Mode profile for heaxapole (219.85 THz) in PhCRR CROW with cavity spacing of 4.320 µm

*Table 2: The normalized group velocity values for PhCRR at the center of the transmission band (calculated using equation 1).*

| CROW | Mode | Lattice constant a, (µm) | Cavity spacing R (µm) | Peak Type | Peak Frequency (ω) THz | f1 (THz) | f2 (THz) | Δf THz | Center frequency (Ω) THz | Coupling coefficient (k) | vg(Ω) m/s | Normalised vg(Ω) |
|---|---|---|---|---|---|---|---|---|---|---|---|---|
| 1 | Hexapole | 0.540 | 3.240 | 1 | 192.28 | 192.27 | 197.193 | 4.92 | 194.73 | 0.01264 | 7.98E+06 | 0.0266 |
| 1 | Hexapole | 0.540 | 3.240 | 2 | 193.91 | 192.27 | 197.193 | 4.92 | 194.73 | 0.01264 | 7.98E+06 | 0.0266 |
| 1 | Hexapole | 0.540 | 3.240 | 3 | 195.78 | 192.27 | 197.193 | 4.92 | 194.73 | 0.01264 | 7.98E+06 | 0.0266 |
| 1 | Hexapole | 0.540 | 3.240 | 4 | 197.18 | 192.27 | 197.193 | 4.92 | 194.73 | 0.01264 | 7.98E+06 | 0.0266 |
| 1 | Octupole | 0.540 | 3.240 | 1 | 219.7 | 219.68 | 221.282 | 1.60 | 220.48 | 0.00363 | 2.60E+06 | 0.0087 |
| 1 | Octupole | 0.540 | 3.240 | 2 | 220.03 | 219.68 | 221.282 | 1.60 | 220.48 | 0.00363 | 2.60E+06 | 0.0087 |
| 1 | Octupole | 0.540 | 3.240 | 3 | 220.66 | 219.68 | 221.282 | 1.60 | 220.48 | 0.00363 | 2.60E+06 | 0.0087 |
| 1 | Octupole | 0.540 | 3.240 | 4 | 221.28 | 219.68 | 221.282 | 1.60 | 220.48 | 0.00363 | 2.60E+06 | 0.0087 |
| 2 | Hexapole | 0.540 | 4.320 | 1 | 194.083 | 194.08 | 194.559 | 0.44 | 194.32 | 0.00112 | 9.42E+05 | 0.0031 |
| 2 | Hexapole | 0.540 | 4.320 | 2 | 194.224 | 194.08 | 194.559 | 0.44 | 194.32 | 0.00112 | 9.42E+05 | 0.0031 |
| 2 | Hexapole | 0.540 | 4.320 | 3 | 194.407 | 194.08 | 194.559 | 0.44 | 194.32 | 0.00112 | 9.42E+05 | 0.0031 |
| 2 | Hexapole | 0.540 | 4.320 | 4 | 194.557 | 194.08 | 194.559 | 0.44 | 194.32 | 0.00112 | 9.42E+05 | 0.0031 |
| 2 | Octupole | 0.540 | 4.320 | 1 | 219.64 | 219.63 | 220.066 | 0.48 | 219.85 | 0.00109 | 1.03E+06 | 0.0034 |
| 2 | Octupole | 0.540 | 4.320 | 2 | 219.76 | 219.63 | 220.066 | 0.48 | 219.85 | 0.00109 | 1.03E+06 | 0.0034 |
| 2 | Octupole | 0.540 | 4.320 | 3 | 219.93 | 219.63 | 220.066 | 0.48 | 219.85 | 0.00109 | 1.03E+06 | 0.0034 |
| 2 | Octupole | 0.540 | 4.320 | 4 | 220.06 | 219.63 | 220.066 | 0.48 | 219.85 | 0.00109 | 1.03E+06 | 0.0034 |

**3 Conclusion**

The MRR CROW was simulated in CST microwave studio. Resonance splitting was observed for resonant modes with even modal numbers. Taking inspiration from this design of MRR CROW a new design for optical delay line based on coupled ring-resonator cavities in photonic crystals is proposed.

A slow-light structure based on four coupled PhCRR cavities was simulated. Multiple resonant modes were observed which demonstrated capability for multiple-channel transfer. Each of the modes had four phase shifted peaks due to delay in each coupled-cavity. To make comparison with MRR CROW delay for heaxapole and octupole was analyzed. For cavity spacing of 3.240 µm: hexapole was centered at 194.73 THz with a normalized group velocity of 0.0266 while octupole was centered at 220.48 THz with a group velocity of 0.0087. When cavity spacing was changed to 4.320 µm: center frequency of hexapole shifted to 194.32 THz and the normalized group velocity changed to 0.0031 while center frequency for octupole shifted to 219.85 THz with a group velocity of 0.0034 . This is comparable to normalized group velocity values 0.0098 and 0.005 for MRR CROW at 196 and 214 THz respectively.

Table 1, 2 and 3 give the coupling coefficients and normalized group velocities for MRR, PhCRR & PhC CROW respectively. MRR CORW provides better reduction in group velocity for comparable coupling coefficients in PhC CROW. While PhCRR gives results similar to MRR CROW but on a smaller length scale therefore highlighting it compactness.

By the changing the cavity spacing the tunabilty of the performance parameters for hexapole and octupole was demonstrated. Additionally, it offers to selectively design performance in the slow-light regime by tuning design parameters such as cavity-size and fill-factor. It was demonstrated that ring-resonator type delay-lines can be executed in PhCs with added advantages over CCWs [10] like multi-channel transfer. The tunabilty of PhCRR based slow-light structure was demonstrated by altering the cavity spacing.

Table 3: Representative coupling coefficients and group velocities (normalised) for CCWs [15]

| CCW mode | Lattice constant a, (µm) | Cavity spacing R (µm) | Coupling coefficient (k) | Normalised vg(Ω) |
|---|---|---|---|---|
| TE | 5.88E-01 | 2a | 4.66E-02 | 0.221 |
| TE | 8.70E-01 | 3a | 1.49E-02 | 0.1 |
| TE | 5.86E-01 | 4a | 4.70E-03 | 0.045 |